\documentclass[prb,twocolumn,floatfix,showpacs]{revtex4}

\usepackage{graphicx}

\usepackage{bm}

\begin{document}


\preprint{}

\title{Fluxoid dynamics in superconducting thin film rings}

\author{J. R. Kirtley and C. C. Tsuei}

\affiliation{ IBM T.J. Watson Research Center, P.O. Box 218, Yorktown
Heights, NY 10598}

\author{V. G. Kogan and J. R. Clem }

\affiliation{ Ames Laboratory and Department of Physics and Astronomy,
Iowa State University, Ames, IA 50011}

\author{H. Raffy and Z. Z. Li}

\affiliation{ Laboratoire de Physique des Solides, Universit{\'e}
Paris-Sud, 91405 Orsay, France}

\date{\today}

\begin{abstract}

We have measured the dynamics of individual magnetic fluxoids entering and
leaving photolithographically patterned thin film rings of the
underdoped
high-temperature superconductor
Bi$_2$Sr$_2$CaCu$_2$O$_{8+\delta}$, using
a variable sample temperature scanning SQUID microscope. These
results can be qualitatively described using a model in which
the
fluxoid number changes by thermally activated nucleation of a
Pearl vortex in, and
transport of the Pearl vortex across, the ring wall.

\end{abstract}


\maketitle



\pagebreak


\section{Introduction}

Although fluxoid quantization in superconductors
was first demonstrated
experimentally over 40 years ago,\cite{deaver,doll}
there has recently been a resurgence of interest in fluxoid
dynamics in a ring geometry.
For example, it has been proposed that the interacting dipole
moments in an array
of superconducting rings can provide a model experimental
system for
studying magnetism in Ising antiferromagnets.
\cite{aeppli,moessner,chandra,davidovic1,davidovic2}
This possibility has become particularly attractive with the
development
of $\pi$-rings: superconducting rings with an intrinsic quantum
mechanical
phase change of $\pi$ upon circling the ring, in the absence of
supercurrents or externally applied fields. Such $\pi$ phase
changes can be
produced either by the momentum
dependence of an unconventional superconducting order parameter,
\cite{geshkenbein,geshkenbein2,sigrice,harlrmp,trirmp} or by
magnetic interactions in the tunneling region of a
Josephson weak link in the
ring.\cite{bulaevski,ryazanov1,ryazanov2}
$\pi$-rings are an ideal model system
for the Ising antiferromagnet, since they
have a degenerate, time-reversed ground state in the absence of
an
externally applied magnetic field. Recent progress
\cite{smildeapl,smildeprl} has
made it possible to reliably make very large closely packed
arrays of
$\pi$-rings, which show strong antiferromagnetic correlations
in their ``spin" orientations upon cooling in zero magnetic
field.\cite{hilgnat}
Superconducting ring experiments
have also been proposed\cite{senthilprl} and performed
\cite{bonn,wynn} to test for the
presence of ``visons", vortex-like topological excitations which may
result from electron fractionalization in the high-T$_c$ cuprate
superconductors. In addition, superconducting rings can
provide a model system for the early evolution of the universe,
through the
study of quenched fluctuations in superconducting rings.
\cite{kibble,kibble2,zurek,carmi1,carmi2,kovoussanaki,ghinovker}
Each of these studies
depend critically on an understanding of the fluxoid dynamics
during the cooldown process.
Finally, there has been a resurgence of interest lately in
macroscopic quantum coherence effects in superconducting
rings, in conjunction with applications to quantum
computation.\cite{mooij,vanderwal,friedman}
In these experiments noise due to thermally
activated motion of fluxoids must be
considered, at least at elevated temperatures. Therefore the study
of fluxoid dynamics in rings is one of much current topical interest.

Thermally activated vortex dynamics in superconductors have been
studied extensively,\cite{blatter} including in
ring geometries.\cite{landau} Magnetic noise in high temperature
cuprate superconductors has been shown to arise from thermally
activated hopping of vortices between pinning sites.\cite{ferrari}
Fluxoid dynamics have been studied in superconducting
rings interrupted by Josephson weak links,\cite{lukens} and in
mesoscopic rings, with the coherence length comparable to the ring
dimensions.\cite{zhang} However,
the system we have chosen to study, relatively large rings with long
magnetic penetration depths (because they are strongly underdoped),
is in many respects
a particularly simple one.
First, our rings have no intentional
Josephson weak links.
Many quantitative
details specific to phenomena related to flux quantization
can be treated within the London approach, which is not bound by the
rigid temperature restriction of the Ginzburg-Landau
theory.
The smallness of the ring with
respect to the Pearl length \cite{pearl} $\Lambda = 2\lambda^2/d$ (in
the thin film
limit, the London penetration depth $\lambda \gg d$, the film
thickness)
simplifies considerably the problem of a ring in an
applied magnetic field.
Studies of the  transitions between quantum states in thin film rings are
relevant for understanding the dynamics  of multiply
connected mesoscopic superconducting devices in general and of
the telegraph noise in these devices in particular.

\section{Experimental results}

\begin{figure}[htb]
  \includegraphics[width=2.5in]{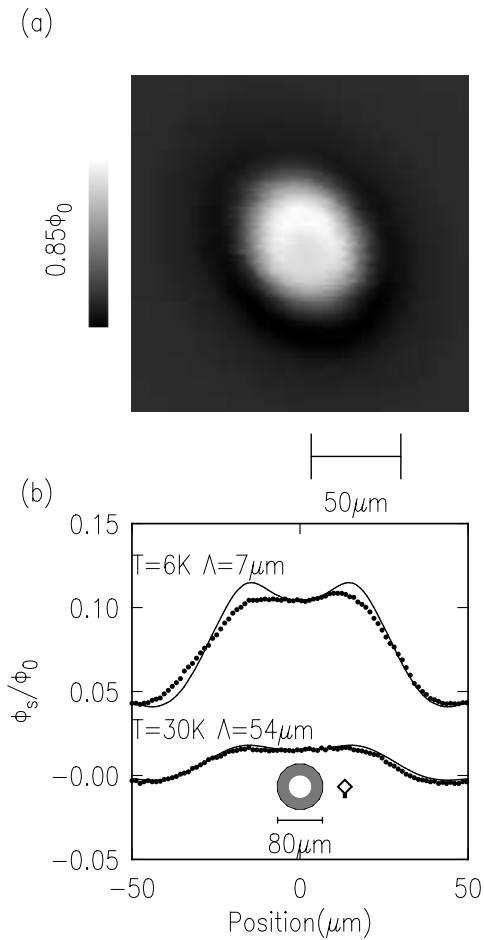}
  \caption{\label{fig:bigcrsvt}(a) Scanning SQUID microscope image of an
  80 {\rm $\mu$m} diameter ring
  cooled in a magnetic induction of 6.9 mG, resulting in a fluxoid number of N = 10,
  and imaged in zero field at T = 6 K.
  (b) Cross-sections through the center of the
  ring in (a), cooled in an induction of 0.7 mG, resulting in a fluxoid
  number of N = 1, and imaged in zero field at temperatures
  of 6 K and 30 K (dots), and modelling
  as described in the text (line). The data and fit for T = 6 K have been
  offset vertically by 0.05 $\phi_0$ for clarity.
  The insets at the bottom of (b) show schematics of the ring
  and SQUID pickup loop geometries.
  $\phi_s$ is the flux through the SQUID pickup loop.
  }
\end{figure}

Our measurements were made on 300 nm thick films of the
high-temperature superconductor
Bi$_2$Sr$_2$CaCu$_2$O$_{8+\delta}$ (BSCCO), epitaxially grown on (100)
SrTiO$_3$
substrates using magnetron sputtering. The oxygen concentration in
these films was
varied by annealing in oxygen or argon at 400-450 $^o$C.
The films were photolithographically
patterned into circular rings using ion etching. The rings had outside
diameters of 40, 60, and 80 {\rm $\mu$m}, with inside diameters half the
outside diameters. The film for the current measurements had a broad
resistive transition (90\% of the extrapolated normal state resistance
at T = 79 K, 10\% at T = 46 K)
with a zero-resistance $T_c$ of 36 K before patterning. Such broad
resistive transitions are characteristic of both single crystals
\cite{watanabe}
and thin films \cite{konstantinovic} of BSCCO, and may be indicative of
oxygen inhomogeneity. In this paper we will treat the
rings as homogeneous and cylindrically symmetric. This view
is supported by two facts: 1) the SQUID images are homogeneous, at
least
within the spatial resolution set by the 17.8 $\mu$m
pickup loop size,
at all temperatures (see e.g. Fig. \ref{fig:bigcrsvt});
and
2) the Pearl penetration depth is quite long, of order 100 $\mu$m, at
the
temperatures of interest. This long penetration depth might be
expected to
average out spatial inhomogeneities. Nevertheless, it is possible that
the fluxoid transitions in our samples are dominated by paths
with relatively low barrier heights.\cite{landau}
It is therefore remarkable that the simple
model described in this paper qualitatively describes our results in
the
presence of this inhomogeneity. The critical temperatures of the
rings were slightly lower after patterning than the blanket coverage
film,
presumably due to additional oxygen removal in the ion
etching step. The critical temperature of
the individual ring being measured was determined
by SQUID inductive
measurements, as described below.
The rings were magnetically imaged using a variable sample temperature
scanning Superconducting Quantum Interference Device (SQUID)
microscope,\cite{vartapl} which scans a sample relative to a
SQUID with a small, well shielded, integrated pickup loop
(a square loop 17.8 $\mu$m on a side for these measurements),
the sample temperature being varied
while the low-$T_c$ SQUID remains superconducting.

\begin{figure}
\includegraphics[width=2.5in]{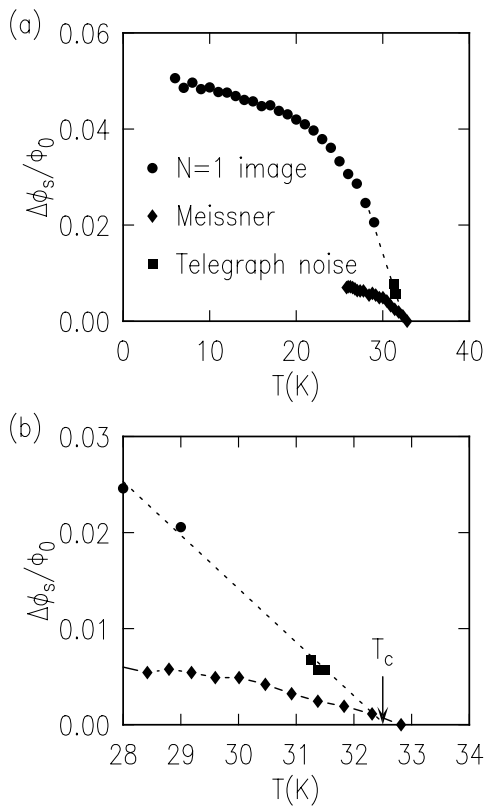}
\caption{\label{fig:bigphivt}(a) Difference $\Delta\phi_s$ in SQUID
signal directly above the 80 $\mu$m
diameter ring minus that with the SQUID far from the ring,
with the ring in the N = 1 fluxoid
state (solid circles); Meissner screening signal $\Delta\phi_s$ with an
applied induction of $\sim$ 0.2 mG (diamonds); and amplitude of the
telegraph
noise due to switching between fluxoid states at $\phi_a = \phi_0/2$ (squares),
all as a function of temperature. (b) Expanded view of the
data close to the ring superconducting temperature $T_c$.
}
\end{figure}

Figure \ref{fig:bigcrsvt}(a)
shows a scanning SQUID microscope image of an 80 $\mu$m outside
diameter
ring, cooled in an induction of 6.9 mG, which results in a vortex number N = 10
in the ring, and imaged in zero field at low temperature.
For consistency, all of the measurements
presented in this paper were made on this ring.
Measurements on a number of rings of all three sizes were also
performed,
with quite comparable results.
The dots in Figure \ref{fig:bigcrsvt}(b) are cross-sections
through the center of the 80 $\mu$m ring with a ring fluxoid number of
N = 1
in zero field at two temperatures. The solid lines are fits
to the SSM images, taking into account the detailed current
distributions
in the rings. Such fits are used in this paper to determine the
temperature-dependent Pearl penetration
length $\Lambda$, which is an important
parameter in modelling the fluxoid dynamics.

\begin{figure}
\includegraphics[width=2.5in]{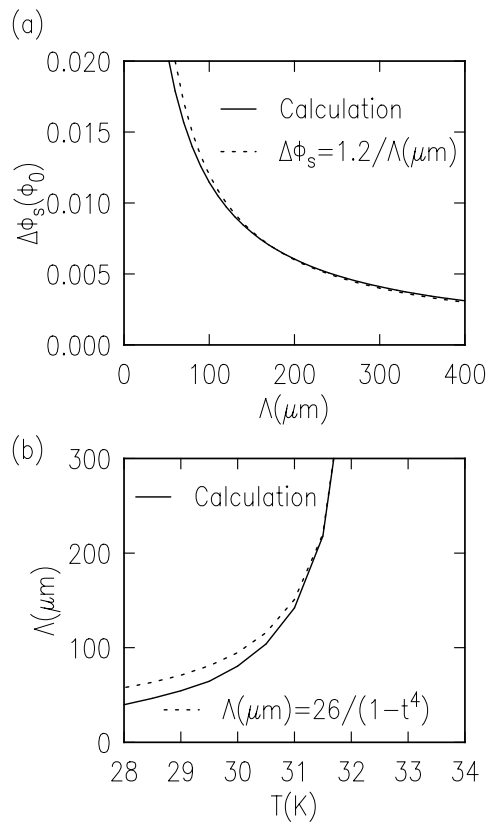}
\caption{\label{fig:biglamvt}(a) Calculated dependence of the SQUID
difference signal
$\Delta\phi_s$ above minus away from the ring, for the ring and pickup
loop geometry used in this paper, as a function of the Pearl length
$\Lambda$ (solid line). The dashed line shows that $\Delta\phi_s$ is
calculated to be nearly inversely proportional to $\Lambda$.
(b) Calculated dependence of the Pearl length $\Lambda$ on temperature
for the 80 $\mu$m ring, assuming the linear dependence of $\Delta\phi_s$
on temperature indicated by the dashed line in Fig. \ref{fig:bigphivt}
(solid line). The dashed line is proportional to $(1-t^4)^{-1}$,
the ideal temperature dependence
for $\Lambda$.}
\end{figure}

Fig. \ref{fig:bigphivt} shows the results of a number of such
measurements as a function of temperature on this ring. The solid
circles
in Fig. \ref{fig:bigphivt}a are the difference $\Delta\phi_s$
between the SQUID signal
with the pickup loop centered over the ring and that with the pickup
loop
far from the ring, with the ring in the N = 1 fluxoid state.
Such measurements cannot be made closer than about 1K from T$_c$
because the
ring switches to the N = 0 state. The solid diamonds are measurements
$\Delta\phi_s$ with a small applied induction $B_a$ = 0.2 mG
with the ring in the N = 0 state. The squares are the change in the flux
through
the SQUID when the ring spontaneously changes fluxoid number in the
telegraph
noise described below. The dashed line is a linear extrapolation of the
$\Delta\phi_s$ (circles and squares) data; the zero crossing of this
line
provides an estimate of the critical temperature T$_c$ for this ring,
T$_c$ = 32.5$\pm$0.2 K.

\begin{figure}
\includegraphics[width=2.5in]{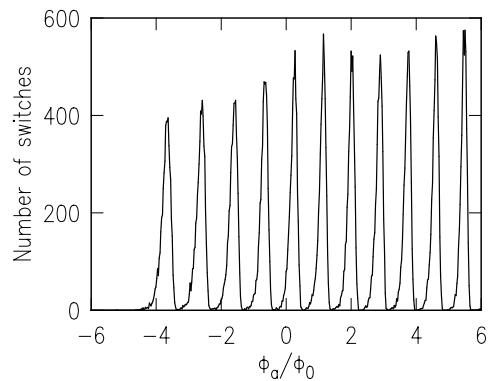}
\caption{\label{fig:histgram}Histogram of the number of switches
observed, as a function
of the externally applied flux, for a SQUID pickup loop positioned
directly above the 80 $\mu$m outer diameter underdoped BSCCO ring at
T = 30.9 K. The sweep
rate was 200$\phi_0$/s, with the data stored in 512 bins.}
\end{figure}

Our modelling of the supercurrent distributions in these rings is as follows:
Consider a thin film ring of thickness $d\ll \lambda$
with radii $a<b$  in the plane $z = 0$. The London equations for the film
interior read

\begin{equation}
\label{eq:London}
{\bf j} = -\frac{c\phi_0}{8\pi^2\lambda^2}\Big(\nabla\theta
   +{2\pi\over\phi_0}\,{\bf A}\Big)\,,
\end{equation}
where ${\bf j}$ is the supercurrent density,
$\phi_0 = hc/2e$ is the superconducting flux quantum,
$\theta$ is the order parameter phase,
and ${\bf A}$
is the vector potential. Since the current in the ring must be single
valued,
$\theta = -\,N\,\varphi\,$,
where $\varphi$ is the azimuth and the integer $N$ is
the winding number (vorticity) of the state.
Integrating ${\bf j}$ over the
film thickness $d$, we obtain:

\begin{equation}
    g_{\varphi}\equiv g(r) = \frac{c\phi_0}{4\pi^2\Lambda}\Big( {N\over r}
   -{2\pi\over\phi_0}\, A_{\varphi}\Big)\,,\label{eq:current}
\end{equation}
   where $g(r)$ is the sheet current density directed along the azimuth
$\varphi$. The vector potential $A_{\varphi}$ can be written as
\begin{equation}
A_{\varphi}(r) = \int_a^b d\rho \,g(\rho) a_{\varphi} (\rho;r,0)
+{r\over 2}\,H\,,
\label{eq:int}
\end{equation}
where the last term represents a uniform applied field $H$ in the $z$
direction and
$a_{\varphi} (\rho;r,z)$ is the vector potential of the field created
by a circular unit current of radius
$\rho$:\cite{LL}
   \begin{eqnarray}
a_{\varphi}
(\rho;r,z) & = &\frac{4}{ck}\sqrt{{\rho\over r}} \Big[\Big(1-{k^2\over
2}\Big){\bm
K}(k) -{\bm E}(k)\Big],\nonumber\\
k^2& = &\frac{4\rho r}{(\rho+r)^2+z^2}\,.\label{eq:A}
\end{eqnarray}
Here, ${\bm K}(k)$ and ${\bm E}(k)$ are the complete elliptic
integrals in the
notation of Ref. \onlinecite{Grad}.

\begin{figure}
\includegraphics[width=2.4in]{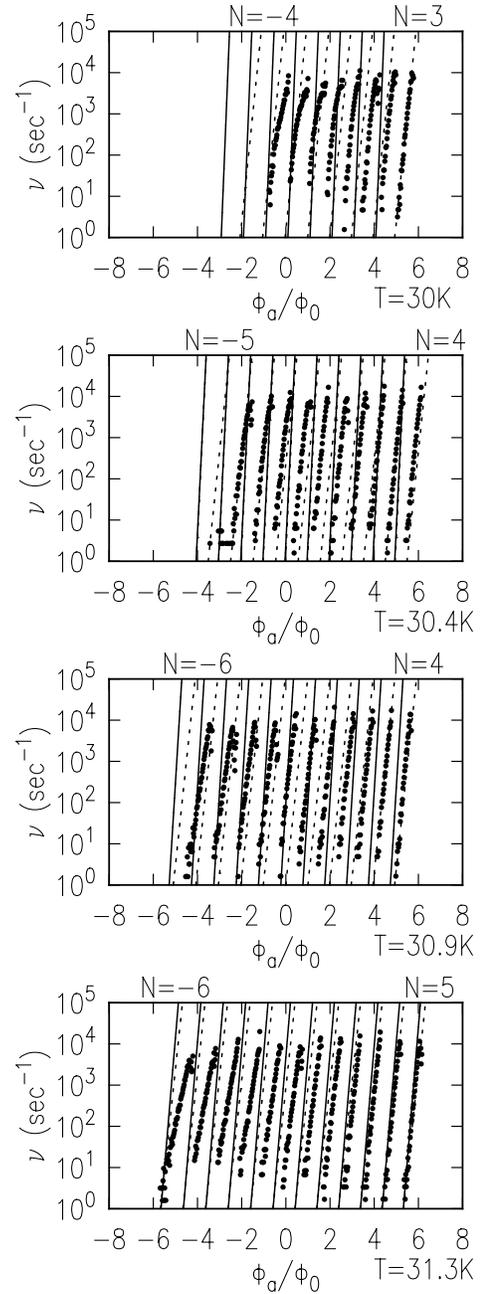}
\caption{\label{fig:ratefig}Fluxoid transition rates $\nu$ for the
transition $N \rightarrow N+1$
vs. the externally applied
flux $\phi_a$ (swept towards positive $\phi_a$)
for a BSCCO ring of 80 $\mu$m outer
diameter, with T$_c$ = 32.5 K, at various temperatures.
The solid symbols are experiment.
The solid and dashed lines are the
predictions of the model described in the text.}
\end{figure}

Substituting Eq. (\ref{eq:int}) and (\ref{eq:A}) into (\ref{eq:current}),
we obtain an integral equation for $g(r)$:
\begin{eqnarray}
\label{eq:giter}
&&\frac{4\pi^2\Lambda}{c}\,r\,g(r)+\pi r^2H-\phi_0N\nonumber\\
&& = -\frac{4\pi}{c}\int_a^b d\rho\,g(\rho)\Big[{\rho^2+r^2\over \rho+r}
{\bm
K}(k_0) -(\rho+r){\bm E}(k_0)\Big],\label{int.eq}
\end{eqnarray}
where $k_0^2 = 4\rho r/(\rho+r)^2$. This equation is solved by iteration
for a given integer $N$ and field $H$ to produce
current  distributions which we label as $g_N(H,r)$.

After $g_N(H,r)$ is found, the field outside the ring can be calculated
using Eq. (\ref{eq:A}):
\begin{eqnarray}
\label{eq:fieldz}
h_z(N;r,z) = \frac{2}{c}\int_a^b\frac{d\rho\,g_N(H,\rho)}{\sqrt{(\rho+r)^2+z^2}}
\Big[ {\bm K}(k) \nonumber\\
+\frac{\rho^2-r^2-z^2}{(\rho-r)^2+z^2} {\bm
E}(k)\Big]+H\,.\label{h(r,z)}
\end{eqnarray}

The flux through the SQUID is obtained numerically by integrating
Eq. (\ref{eq:fieldz}) over the pickup-loop area. The lines in Fig.
\ref{fig:bigcrsvt}b are
two-parameter fits of this integration of
Eq. (\ref{eq:fieldz}) to the data, resulting in
$z$ = 3.5 $\mu$m, and $\Lambda$ = 7 $\mu$m (corresponding to
$\lambda$ = 1 $\mu$m) at T = 6 K, and $\Lambda$ = 54 $\mu$m
($\lambda$ = 2.8 $\mu$m) at T = 30 K.

This value ($\lambda$ = 1$\mu$m) for the low-temperature
in-plane penetration depth
is at first surprising, given the observed values of $\lambda \approx$ 0.2
$\mu$m for BSCCO near optimal doping.\cite{waldmann} However, one might
expect the penetration depth to be larger for our underdoped films because
of their lower T$_c$, following the Uemura relation
$\lambda^{-2} \sim$ T$_c$.~\cite{uemura} Further, these films have
large normal-state resistivities
$\rho \approx$ 1200 $\mu\Omega$-cm, meaning that they are in the dirty limit,
and close to the metal-insulator transition.\cite{semba} The zero-temperature
penetration
depth of a dirty-limit superconductor is given by  $\lambda_0 = (c/2\pi)\sqrt{\hbar\rho/\Delta_0}$.
Taking the BCS value $\Delta_0 = 1.74 k_B T_c$, with T$_c$ = 30 K gives
$\lambda_0$ = 0.7 $\mu$m. It is expected that fluctuations in the
superfluid density could further increase the penetration depth in these
layered superconductors.\cite{artemenko}

To model the fluxoid dynamics data presented in this paper,
it is necessary to estimate the
temperature-dependent Pearl length $\Lambda$ and the energy
associated with supercurrent flow
in our rings.
We can infer the temperature dependence of the Pearl length from the
temperature dependence of $\Delta\phi_s$ as follows. Numerical
integration
of Eq. (\ref{eq:fieldz}) for our ring and SQUID pickup loop geometry as
a function of the Pearl length $\Lambda$ gives
the solid line in Fig. \ref{fig:biglamvt}a. The calculated
$\Delta\phi_s$
is nearly inversely proportional to $\Lambda$, as shown by the dashed
line in Fig. \ref{fig:biglamvt}a. The linear dependence
of $\Delta\phi_s$ on temperature indicated by the dashed line in
Fig. \ref{fig:bigphivt} results in a temperature
dependence of the Pearl length $\Lambda(T)$
for this ring indicated by the solid line in Fig. \ref{fig:biglamvt}.
Since the London penetration depth $\lambda \propto 1/\sqrt{1-t^4}$
($t$ = T/T$_c$),
\cite{tinkham}
$\Lambda = 2\lambda^2/d$ should be approximately proportional to
$(1-t^4)^{-1}$ as
indicated by the dashed line in Fig. \ref{fig:biglamvt}b.

\begin{figure}
\includegraphics[width=2.5in]{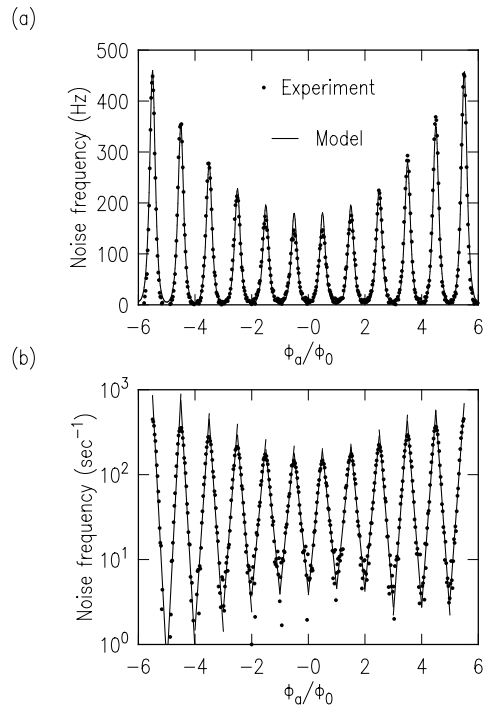}
\caption{\label{fig:teltoyft}(a)Telegraph noise signal vs. applied
flux $\phi_a$ for the
80 $\mu$m ring at T = 31.6 K. The dots are the
data, the solid line is a fit to the model described in the
text.
(b) Replot of the data of (a) on a log-linear
scale. The dots are the data, the solid lines are fits
to an exponential dependence on $\phi_a$, in
segments $N\phi_0 < \phi_a < (N+1/2)\phi_0$ and $(N+1/2)\phi_0 <
\phi_a < N\phi_0$, $N$ an integer.
}
\end{figure}

The fluxoid number $N$ of a ring can be changed by varying the externally
applied flux $\phi_a = H A_{eff}$, where $A_{eff}$ is the effective area,~\cite{clemarea}
and can be monitored by positioning the SQUID pickup loop directly over it.~ \cite{triprl}
In the limit $\Lambda >>b$, the current around the loop can be found by integrating
Eq. (2) to obtain $A_{eff} = (\pi/2) (b^2-a^2)/\ln(b/a)$.
This result also can be obtained from more detailed calculations of
the system energy E(N,H) in equilibrium.~\cite{kogmint}
We assign an experimental value for the effective ring area using the
telegraph noise data of Fig. \ref{fig:teltoyft}, and assuming the peaks are
spaced by $\phi_0$. This gives a value of 2895 $\mu$m$^2$, in comparison with
the calculated value of 2719 $\mu$m$^2$. This discrepancy of about 6\% could
be due to variations in the photolithography of the rings, or errors in the
calibration of the Helmholtz coils which apply the magnetic fields. We used
the experimental value for $A_{eff}$ in determining the flux scales in
Figs. \ref{fig:histgram}, \ref{fig:ratefig}, and \ref{fig:teltoyft}.
At temperatures sufficiently close to T$_c$ the fluxoid
number changes by one fluxoid at a time,
as determined by the agreement (to within 10\%)
with our calculations for $\mid \Delta N \mid = 1$
of the measured spacing in applied flux
between vortex switching events.
Switching distributions $P(\phi_{a,i})$ were obtained by repeatedly
sweeping the applied field,
in analogy with measurements of Josephson
junctions switching into the voltage state.\cite{fulton}
An example is shown in Fig. \ref{fig:histgram}

The transition rates $\nu$
of the fluxoid states were determined from such data using\cite{fulton}
\begin{equation}
\nu(\phi_{a,m}) = \frac{d\phi_a/dt}{\Delta \phi_a}\ln \left \{
\sum^{m}_{j=1}P(\phi_{a,j}) \left /
\sum^{m-1}_{i=1}P(\phi_{a,i}) \right .  \right \},
\label{eq:fulton}
\end{equation}
where $m = 1$ labels the largest $\phi_{a}$
in a given switching histogram peak, and $\Delta\phi_{a}$ is
the flux interval between data points.
The dots in Figure \ref{fig:ratefig} show the results from such
experiments
from the 80 $\mu$m ring at several reduced temperatures.
The assignment of the starting fluxoid number $N$ in this data was made
by following the transitions as they evolved with temperature from the
two-state telegraph noise (see Fig. \ref{fig:teltoyft}) described
below.

\begin{figure}
\includegraphics[width=2.5in]{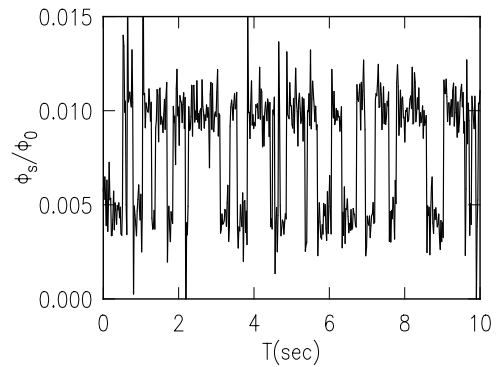}
\caption{\label{fig:teltime}Telegraph noise signal vs. time for the
ring of Figure \ref{fig:histgram} at T = 31.4 K, $\phi_a = \phi_0/2$.
}
\end{figure}

At temperatures sufficiently close to T$_c$ and applied fluxes close to
a half-integer multiple of $\phi_0$, two-state telegraph noise was
observed
in the SQUID pickup loop signal when the loop was placed directly
above a ring.
An example is shown in Fig. \ref{fig:teltime}.

The frequency of this telegraph noise oscillates with the applied
flux, with period $\phi_0$, and peaks at $\phi_a = (N+1/2)\phi_0$, $N$ an
integer,
as shown in Fig. \ref{fig:teltoyft}.

\section{Discussion}

Several general observations can be made about the fluxoid
dynamics observed in our experiments. First, the dynamics
are nearly periodic in the applied field, with a period given
by the applied field times the effective ring area
$A_{eff} = (\pi/2) (b^2-a^2)/\ln(b/a)$
(see
Figs. \ref{fig:histgram}, \ref{fig:ratefig},
and \ref{fig:teltoyft}). This scaling with the effective ring
area has been confirmed for three different ring sizes.

Second, the fluxoid transition rates
depend exponentially on the applied flux,  both for the
fluxoid escape measurements of Fig. \ref{fig:ratefig},
and in the telegraph noise data of Fig. \ref{fig:teltoyft}. The latter
becomes clear when this data is plotted on a log-linear scale,
as in Fig. \ref{fig:teltoyft}(b).

\begin{figure}
\includegraphics[width=2.5in]{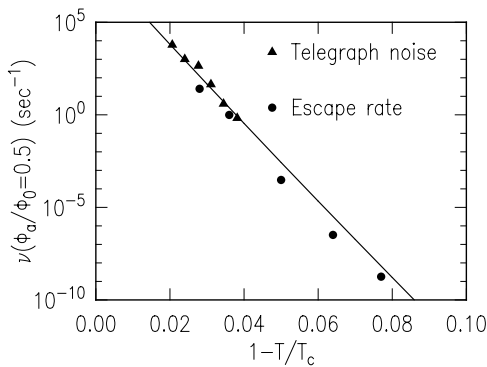}
\caption{\label{fig:ratevst2}Plots of the telegraph noise frequency
and the fluxoid transition rate, for the N = 0 to N = 1
transition, at an applied flux
of $\phi_a = \phi_0/2$, as a function of temperature. The symbols
are the data; the solid line is the model described
in the text. The fluxoid transition rate data was obtained
by extrapolating straight line fits to the data of Fig. \ref{fig:ratefig}.}
\end{figure}

Third, at a particular applied field, both the fluxoid escape
rates and the telegraph noise frequencies depend exponentially
on temperature. An example for the 80 $\mu$m ring
is shown in Fig. \ref{fig:ratevst2}.

We consider the mechanism for transitions between fluxoid
states as a thermally activated nucleation of a vortex
in, and transport of this vortex across, the ring wall. The
relevant energies in the proposed process are 1) the energy required
to nucleate a  vortex, and 2) the kinetic and magnetic energies
associated with supercurrents
in the ring.

The
maximum vortex energy in a straight thin film
superconducting strip
of width $W \ll \Lambda$ (carrying no transport
supercurrent):\cite{kogan94}
\begin{equation}
E_v = \frac{\phi_0^2}{8\pi^2\Lambda}\,\ln\frac{2W}{\pi\xi} \,.\label{eq8}
\end{equation}
where $\xi$  is the vortex core size.
The energy of the ring in a state with the winding number $N$
is\cite{barone}
  \begin{equation}
E_r(N,H)  = E_0\,(N-\phi_a/\phi_0)^2 \,.\label{eq9}
\end{equation}
Clearly, the prefactor $E_0$ coincides
with the ring energy in the state $N = 1$ in   zero applied field:
  \begin{equation}
E_0 = E_r(1,0)  = \frac{\phi_0}{2c} \int_a^b g_{N = 1}(0,r)\,dr\,;
  \label{eq10}
\end{equation}
see Appendix A.

\begin{figure}
\includegraphics[width=2.5in]{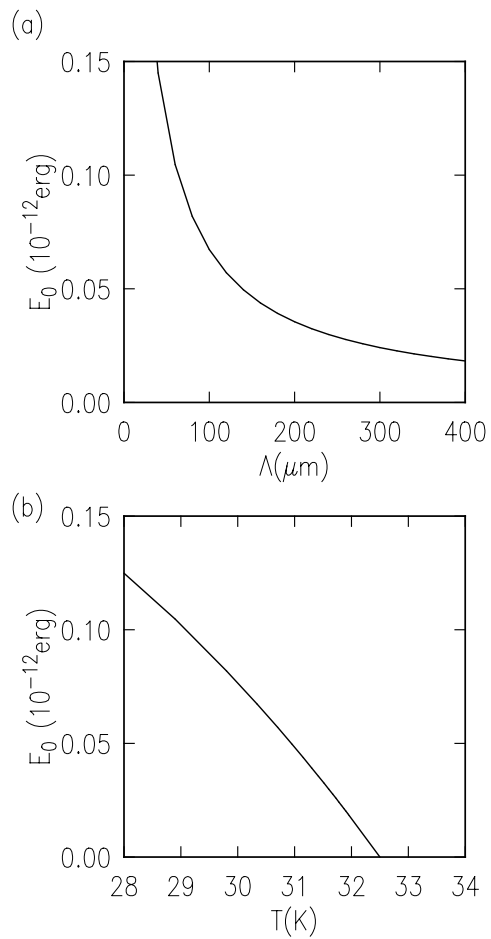}
\caption{\label{fig:bigengvt}(a) Calculated dependence of the total
energy of
our 80 $\mu$m outside diameter ring for $N = 1$, $H = 0$
on the Pearl length $\Lambda$.
(b) Calculated temperature dependence of the total energy for $N = 1$,
$H = 0$,
using the calculated
temperature dependence (solid line) of $\Lambda$ from Fig.
\ref{fig:biglamvt}. }
\end{figure}

We inferred the temperature dependence of the Pearl length from our
SQUID
microscope measurements above (see Fig. \ref{fig:biglamvt}).
Once the Pearl length is known, it is possible to calculate the
temperature
dependence of the energy of our ring. This is done by setting
$N = 1$ and $H = 0$, and integrating the solution of Eq. (\ref{eq:giter})
to obtain
the total supercurrent, and the total energy in the ring from
Eq. (\ref{eq10}).
Figure \ref{fig:bigengvt}a shows the results
of such a calculation for $E_{0}$ as a function of $\Lambda$. Figure
\ref{fig:bigengvt}b
plots $E_{0}$ as a function of $T$ for the $80\, \mu$m ring.

\begin{figure}
\includegraphics[width=2.5in]{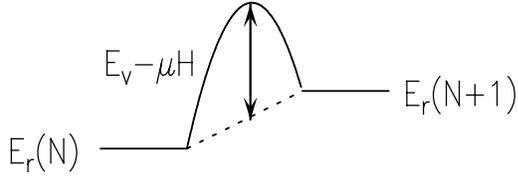}
\caption{\label{fig:ringedia}Schematic energy level diagram
for the thermally activated vortex transport mechanism for fluxoid
jumps proposed in this paper}
\end{figure}

Figure \ref{fig:ringedia} shows a simplified schematic of the
energies involved  in the thermally activated process $N\to N+1$ which
is accomplished by a vortex (or an antivortex) crossing the ring.
The ring has an initial ring energy $E_r(N)$, and a final
ring energy $E_r(N+1)$. Within this simple scheme, the energy barrier
for the process is
  \begin{eqnarray}
\Delta E  = E_v &\mp& \mu
H+\frac{E_r(N+1)+E_r(N)}{2}-E_r(N)\nonumber\\
= E_v &\mp& \mu H+\frac{E_r(N+1)-E_r(N)}{2}\nonumber\\
  = E_v &\mp& \mu H+
E_0(N-\phi_a/\phi_0+1/2)\,,
\label{eq11}
\end{eqnarray}
Here, we have used Eq. (\ref{eq9}); the upper (lower) sign is for a vortex
(antivortex) since the corresponding energy is $-{\bm\mu}\cdot{\bm H}$.

It should be noted that the model we consider here is by no
means exact. It disregards an intricate interplay between   vortex
currents   and those flowing in the ring in a certain quantum
state $N$ (the currents in the ring are not a simple superposition of
vortex currents and those in the absence of a vortex - even within a linear
London approach - because a vortex causes the vorticity $N$ to depend on
the vortex position).
For this reason, there is no point -
within our model - to calculate ``exactly"  the magnetic moment; instead,
we consider $\mu$ as a fitting parameter.

We further simplify the model by considering only  transitions between
the ground state and the first excited state as in the case of a two-level
system.  For the two-level system, the RTN rate
$\nu = P_1/\tau_1 = P_2/\tau_2$, where
$P_{1,2}$ are probabilities to find the system in the states 1,2 and
$\tau_{1,2}$ are the lifetimes. Since $P_1+P_2 = 1$, we readily get
$P_{1,2} = \tau_{1,2}/(\tau_1+\tau_2)$,\cite{machlup} and
\begin{equation}
\nu = {1\over \tau_1+\tau_2}\,. \label{nu}
\end{equation}

If the system is in the ground state $N$, the closest state of a higher
energy depends on the applied field. Using Eq. (\ref{eq9}) it easy to
verify that for $N-1/2<\phi_a/\phi_0<N$, the closest state is $N-1$, whereas
for $N<\phi_a/\phi_0<N+1/2$, the first excited state is $N+1$. We begin with
the latter possibility. The rate of the transition $N\to N+1$ is
\begin{equation}
\tau^{-1}_{N,N+1} =
\nu_0\,\big(e^{-U_v(N,N+1)}+ \,e^{-U_{av}(N,N+1)}\big) \label{N,N+1}
\end{equation}
since the transition can be accomplished by both vortices and
antivortices. Here, $\nu_0$ is an``attempt
frequency",  $U_v$ and $U_{av}$ denote  corresponding barriers
divided by $k_BT$ for vortices and antivortices (for brevity the argument
$\phi_a/\phi_0$ of the $U$'s is omitted).  This expression can be easily factorized
with the help of an identity $e^x+e^y = 2\cosh[(x-y)/2]\,\exp[(x+y)/2]$:
\begin{equation}
\tau_{N,N+1} = \frac{\exp\{[E_v+E_0(N-\phi_a/\phi_0+1/2)]/T\}}
{2\nu_0\,\cosh(\mu H/ T)}\,,
\label{t1}
\end{equation}
where Eq. (\ref{eq11}) has been used and for brevity we set $k_B = 1$.
Similarly we obtain:
\begin{equation}
\tau_{N+1,N} = \frac{\exp\{[E_v-E_0(N-\phi_a/\phi_0+1/2)]/T\}}
{2\nu_0\,\cosh(\mu H/ T)}\,.
\label{t2}
\end{equation}
Now Eq. (\ref{nu}) yields
\begin{equation}
\nu = \nu_0\,e^{-E_v/T}\frac{\cosh(\mu H/
T)}{\cosh[E_0(N-\phi_a/\phi_0+1/2)/T] }\,.
\label{nu1}
\end{equation}
The same calculation for the applied field $N-1/2<\phi_a/\phi_0<N$ gives
\begin{equation}
\nu = \nu_0\,e^{-E_v/T}\frac{\cosh(\mu H/
T)}{\cosh[E_0(N-\phi_a/\phi_0-1/2)/T] }\,.
\label{nu2}
\end{equation}

The factors $1/\cosh[E_0(N-\phi_a/\phi_0\pm 1/2)/T] $    oscillate with
the period $\Delta \phi_a/\phi_0 = 1$ because in the ground state the number
$N$ is the closest integer to the value of $\phi_a/\phi_0$. Due to these
factors,
$\nu(\phi_a/\phi_0)$ has maxima at $\phi_a/\phi_0 = N\pm 1/2$. Clearly, the peaks of
$\nu(\phi_a/\phi_0)$ become sharper when the parameter $E_0/T$ increases.

The numerator $\cosh(\mu H/T)$ provides an increase of the maxima
with increasing applied field.  Physically, this happens because the
vortex magnetic moment reduces the energy barrier by $\mu H$.
If $\mu H/T\ll 1$, the maxima increase quadratically with field:
$\cosh(\mu H/T)\approx 1+ \mu^2 H^2/2T^2$. This is, in fact, the case
for our data. One does not expect $\nu(\phi_a/\phi_0)$ to increase without a
limit: at a certain applied field, the barrier for the vortex entry
splits in two and the vortex can stay in a metastable equilibrium at
the ring. Our model does not hold for such fields.

The solid line in Fig. \ref{fig:teltoyft} shows a fit of Eqs.
(\ref{nu1}) and (\ref{nu2})
to the experimental data. The best fit parameters were
$\nu_0$ = 1.1$\times$10$^8$ s$^{-1}$,
$E_v = 6.03\times 10^{-14}\,$erg,
$E_{0} = 4.98\times 10^{-14}\,$erg, and
$\mu = 1.93\times 10^{-13}\,$erg/G.
 From Fig. \ref{fig:biglamvt}b we read $\Lambda$(T = 31.6 K) = 240 $\mu$m.
Taking
$\xi = 3.2/\sqrt{1-t}$ nm,\cite{seidell}
and $W = 20 \mu$m, we calculate
$E_{v} = 1.46\times 10^{-13}$ erg,
a factor of 2.4 larger than the value extracted from the fit. As
discussed above, sample inhomogeneities or surface
defects\cite{vodolazov} could reduce
the barrier to entry of vortices in type-II superconductors.
 From Fig. \ref{fig:bigengvt}b we read
$E_{0} = 2.2\times 10^{-14}\,$erg,
smaller than the value obtained from the fit by a factor of
2.3. Our value for the attempt
frequency is within the range (10$^6$-10$^{10}$ Hz) suggested
for attempt frequencies from experiments on thermally
activated flux jumps in high temperature superconductors.
\cite{crisan,yeshurun}
Therefore our model provides
a good description of the magnetic field dependence of the telegraph
noise
at a fixed temperature, using values for the vortex nucleation energy
$E_{v}$ and the ring supercurrent energy coefficient $E_{0}$ that are
within approximately factors of 2 of values calculated from
experimental measurements on the same ring.

Note that our estimate of the attempt frequency  is very sensitive to the
value of the coherence length $\xi$. Indeed, the factor $\exp(-E_v/T)$
combined with $E_v$ of Eq. (\ref{eq8}) yields
\begin{equation}
\nu \propto \Big({\pi \xi\over2W}\Big)^{\phi_0^2/8\pi^2\Lambda T} \label{xi^b}
\end{equation}
with a large exponent $\phi_0^2/8\pi^2\Lambda T$.

The same model provides good
agreement with the temperature dependence of the fluxoid transition
rates and telegraph noise frequencies for the $N = 0\to N = 1$
transition at $\phi_a = \phi_0/2$
shown in Fig. \ref{fig:ratevst2}. The solid line in Fig.
\ref{fig:ratevst2}
is the prediction of Eq. (\ref{N,N+1}), scaling the value
for
$E_{v}(T)$ at T = 31.6 K from the fit of Fig. \ref{fig:teltoyft} by
$E_{v}(T) = E_{v}(T = 0)(1-t^4)$.

The solid lines in Fig. \ref{fig:ratefig} show the predictions of
Eq. (\ref{t1}), using the model outlined above,
with the fit values from the telegraph noise data of Fig.
\ref{fig:teltoyft},
with $E_{v}$ and $E_{0}$ scaled in temperature according to the
calculated curves
in Figs. \ref{fig:biglamvt} and \ref{fig:bigengvt} respectively. The
predictions
of the model diverge from experiment for lower temperatures and
fluxoid numbers. In particular, the model predicts
that the slope of the fluxoid transition rates with applied flux
should {\it increase} as the temperature is reduced. However, as can
be seen
from Fig. \ref{fig:ratefig}, although these slopes are
relatively insensitive to temperature, if anything they {\it decrease}
with
decreasing temperature. Somewhat better agreement with experiment
(the dashed lines in Fig. \ref{fig:ratefig}) is obtained
if $E_{0}$ is taken to have the temperature independent value
obtained from
the fit to telegraph noise data of Fig. \ref{fig:teltoyft}, with
$E_v(T)
= E_v(T = 0)(1-t^4)$ as before.

We can speculate on some of the sources of the differences between
the predictions of our model and experiment. First, the
model does not take into account interactions between the bulk vortex
and the supercurrents. As discussed above, we have also implicitly
assumed that the rings
are spatially homogeneous, with a sharp superconducting transition
temperature.
The resistive transitions are in fact quite broad. This broadening
could
be a source of the apparently reduced temperature dependence
of $E_{0}$ and reduced vortex nucleation energy that we observe.
Finally, spatial inhomogeneities could reduce the effective
width of the rings.

In summary then, we have measured single fluxoid transitions and two-state
telegraph noise in superconducting thin film rings as a function of
applied
magnetic field and temperature at temperatures close
to T$_c$. The long penetration depths in the underdoped cuprate
films used allowed measurements over a relatively broad temperature
range.
The measurements are generally consistent with a model in which the
fluxoid transitions are mediated by thermally activated nucleation of a
bulk vortex in, and transport of the vortex across, the ring wall. We
presented a model which qualitatively
explains some of the features of the data, but other features remain
puzzling.

We would like to thank T. Senthil and M.P.A. Fisher for suggesting ring
experiments to us, and
D. Bonn,
W.A. Hardy,
R. Koch,
K.A. Moler,
D.J. Scalapino, F. Tafuri,
and
S. Woods
for useful conversations. We would like to thank R. Mints for detailed
discussions and calculations of the fluxoid transition process.
This manuscript has been authored in part by
Iowa State University of Science and Technology under Contract
No. W-7405-ENG-82 with the U.S. Department of Energy.
The work of VGK was partially supported by the
Binational US-Israel Science Foundation.

\appendix
\section{}

     The magnetic part of the energy for the state $N$ in zero applied field is
  $E_m = \int d^2{\bm r} \,\,{\bm A}\cdot{\bm g}/2c$. Substitute here
the vector potential from Eq. (2) to obtain $E_m = -(\pi
\Lambda/c^2)\int  d^2{\bm r}\,g^2 - (\phi_0/4\pi
c)\int d^2 {\bm r}\,\nabla\theta \cdot{\bm g}$. Since the kinetic part is
the integral over the volume of the quantity
$2\pi\lambda_L^2\,j^2/c^2 = \pi\Lambda g^2/c^2d$, the first term in $E_m$
is $-E_{kin} $. Further,
$\nabla_{\varphi}\theta = -N/r$ and we have $E_m+E_{kin} = (\phi_0N/2 c)\int
dr \,g_{\varphi}(0,r)$.




\end{document}